\documentstyle[citesort,pre,aps,multicol,graphicx]{revtex}

\begin{document}
\draft
\title{Experimental and numerical signatures 
of dynamical crossover in orientationally
disordered crystals}
\author{F.~Affouard, E.~Cochin, R.~Decressain and M.~Descamps}
\address{
Laboratoire de Dynamique et Structure des Mat\'eriaux Mol\'eculaires,\\
CNRS ESA 8024,
Universit\'e Lille I,\\
 59655 Villeneuve d'Ascq Cedex France}
\date{\today}
\maketitle

\begin{abstract}
By means of NMR experiment and MD computer simulation 
we investigate the dynamical properties of a
chloroadamantane orientationally
disordered crystal. 
We find a plastic-plastic dynamical
transition at $T_{x} \simeq 330 $ K in the pico-nanosecond regime.
It is interpreted 
as the rotational 
analogue of the Goldstein
crossing temperature between quasi-free diffusion and activated regime
predicted in liquids. Below $T_{x}$, NMR experimental 
data are well described by a Frenkel model corresponding to 
a strongly anisotropic motion. At higher temperatures, 
a drastic deviation is observed 
toward quasi-isotropic rotational diffusion.
Close to $T_{x}$, we observe that two-step relaxations
emerge.
An interpretation which is based on the present study
of a specific heat anomaly detected
by a recent calorimetric experiment is proposed.
\end{abstract}

\begin{multicols}{2}
\narrowtext

The elusive nature of the glass formation relates to the
sharp rise of the transport coefficients (viscosity or
relaxation times) 
in a narrow temperature range above
the calorimetric transition temperature $T_{g}$~\cite{Ediger}.
Some of the recent active investigations focus on 
the existence of a dynamical crossover 
which could occur above $T_{g}$ in glass-forming liquids. The
idealized version of the Mode Coupling Theory(MCT)~\cite{Gotze} 
predicts 
that a dynamical decoupling
occurs at a critical temperature $T_{c} > T_{g}$~\cite{Rossler_prl}.
It is however suspected that $T_{c}$ corresponds to or is close to
the temperature $T_{x}$ at which the system also starts
being sensitive to
the energy landscape in its exploration of the configurational
hyperspace~\cite{Sastry,Glotzer,Heuer}.
Motions in liquids would thus be
dominate at low temperature by activated processes
between potential energy minima
while free diffusion occurs at high temperatures.
Such dynamical crossover was already argued by Goldstein
thirty years ago~\cite{Goldstein}.
In the last few years a significant effort has been undertaken to 
investigate experimentally and numerically 
the properties of glass-forming liquids in order to detect 
such dynamical changeover. 
However, most of the good glass-formers 
which allow easy investigations of the supercooled metastable state
such as o-terphenyl~\cite{Lewis2}
are molecular liquids 
where contributions due to
translational, orientational and internal degrees of freedom 
(TDOF, ODOF) 
are inextricably mixed. It makes the properties 
of the energy landscape topography, MCT predictions
or detection of dynamical transitions
extremely hard to investigate.
A clear visualization of the different processes 
at the molecular level of simple model compounds is needed.
The discovery that some orientationally disordered crystals 
exhibit the phenomenology of conventional glasses is therefore
an exciting opportunity.

Indeed, some molecular crystals 
present 
a \emph{plastic phase} 
in which the molecular centers of mass 
show a perfect average crystalline translational order
while the orientations are
dynamically disordered.
Some plastic crystals, {\em i.e}
glassy crystals~\cite{Suga,Descamps4}, can be undercooled 
and present many properties characteristic of
conventional molecular-liquid glasses,
both in the way the
glass occurs (undercooling)
and the thermodynamic signatures.
Glassy crystals which are the very rotational analogue of liquid
glass-formers
offer an excellent opportunity to focus on the role of the ODOF.
In~\cite{Renner}, Barrat \emph{et Al.} investigated the rotational
dynamics of a colliding hard needle model on a rigid lattice
and have shown the emergence of a dynamical decoupling.
Substituted adamantanes plastic crystals
such as the cyanoadamantane~\cite{Descamps3}
noted CNa in the following
are good experimental candidates and 
some of them
exhibit a glass transition signature.
In~\cite{Affouard}, MD numerical simulations of 
a simple model inspired from the CNa molecular geometry
allowed us to investigate both TDOF and ODOF properties. In this work
a dynamical decoupling and two-step relaxations 
were found to be possible in a rotator phase.

\begin{figure}[h]
\includegraphics[width=6.5cm]{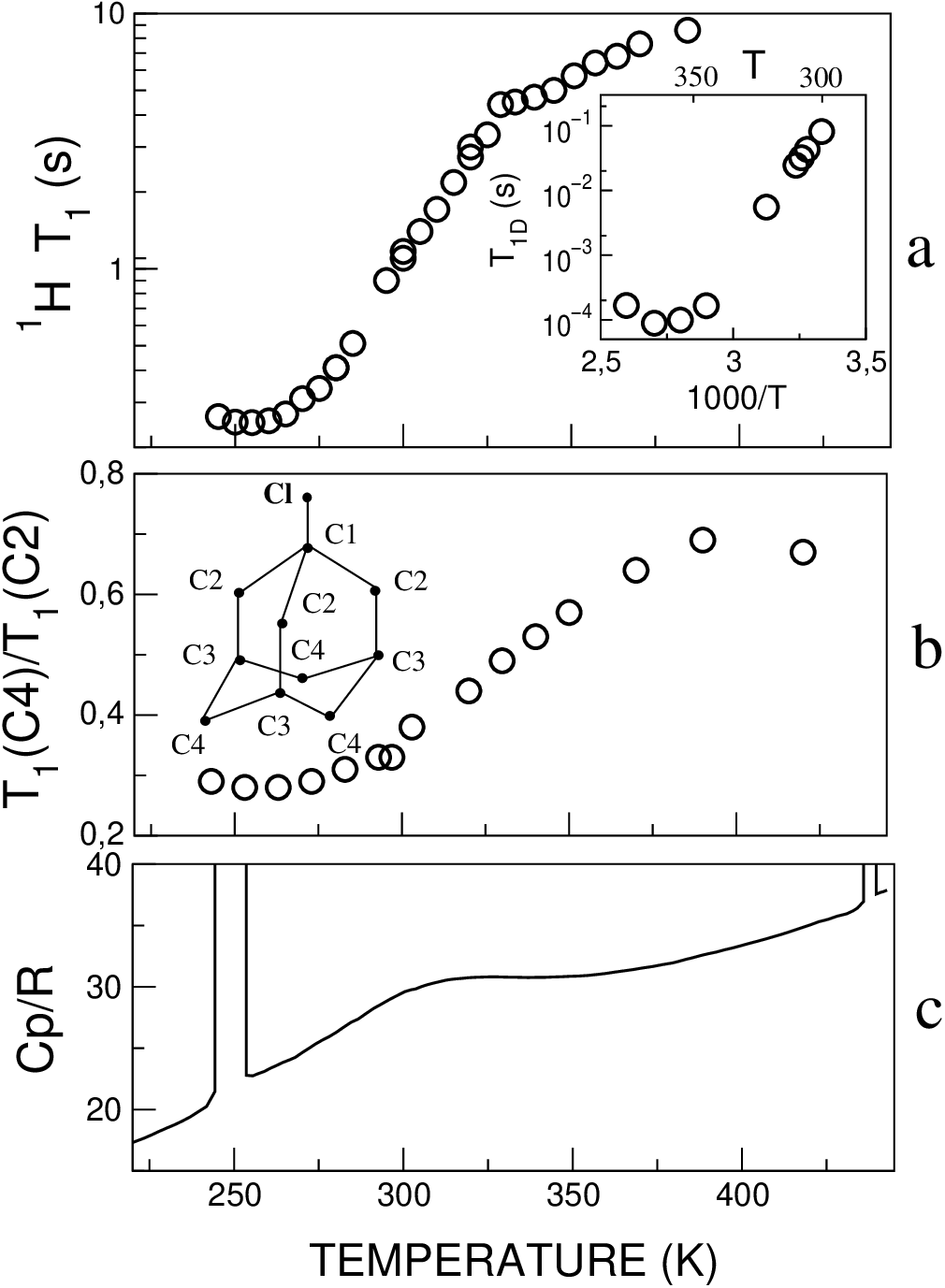}
\caption{\protect\footnotesize Experimental data as
function of temperature in Cla plastic phase.
{\bf a},
the experimental proton relaxation time $T_{1}$ and 
the dipolar order $T_{1D}$.
{\bf b}, the ratio $T_{1}(C4)/T_{1}(C2)$ 
of the methylene relaxation times. 
The chloroadamantane molecule with the different carbons
is displayed in the inset.
{\bf c}, Experimental heat capacity where
data have been
extracted from Oguni et Al (Ref~\protect\cite{Oguni}).
}
\label{figure2}
\end{figure}

In the present letter, we address the possibility of finding such a behaviour in
a real system. 
We have found chloroadamantane $\mathrm{C_{10}H_{15}Cl}$ (Cla) to be a very favorable
and relevant system for this investigation. It 
shows a plastic phase
over a wide interval of temperature $[244-442]$ K (see Fig.~\ref{figure2}c)
and a structure isomorphous
to CNa~\cite{Foulon}.
The Cla molecule possesses
a smaller
substitute than does CNa. This gives rise to 
faster dynamics in the plastic phase
where rotational motions are also suspected to change in nature as 
it is reported from earlier incoherent
quasielastic neutron scattering experiments~\cite{Bee}.
Our results show for the first time, for a rotator phase,
using experimental NMR and numerical MD simulations,
the evidence
of a dynamical crossover between quasi-free rotational diffusion 
and activated motion occurring in the pico-nanosecond regime.
It is also shown that this
accident can explain the existence of a 
calorimetric anomaly (see Fig.~\ref{figure2}c)
recently detected in Cla by Oguni \emph{et Al.}~\cite{Oguni}.

NMR experiments were performed on Chloroadamantane on Bruker spectrometers
ASX100 and AMX 400. 
$\mathrm{^{1}H}$ and $\mathrm{^{13}C}$ linewidths and relaxation times were measured. 
The experiments were carried out on powder samples sealed in glass tubes
 for $\mathrm{^{1}H}$ 
experiments or packed into zirconium rotors for  $\mathrm{^{13}C}$ MAS experiments. 
The spinning speed of the sample was 4 kHz. 
The sample temperature was regulated to within $\pm 0.5$ K by a Brucker BVT 2000 
temperature controller.
The spin lattice relaxation times were measured using the inversion recovery pulse sequence,
the recovery of the magnetization was always found to be exponential
within experimental accuracy. The $\mathrm{^{1}H}$ spin lattice $T_{1}$ vs temperature
is reported in Fig.~\ref{figure2}a. The main feature of this curve
is the occurrence 
of an abrupt change of the activation energy, from 1400 K to 3400 K,
at a temperature close to $T_{x} \simeq 330 $ K
which is far from both the melting temperature and the brittle to plastic transition.
Slow diffusive motions of the molecules
are generally observed in plastic phases.
In the present study, their contribution to the spin-lattice relaxation
time $T_{1}$ was found to be negligible in the temperature range investigated.
This fact was particularly evidenced by measurements of 
the dipolar order $T_{1D}$ (see Fig.~\ref{figure2}a).
The diffusional correlation time ($\tau_{diff} \simeq 30000$ s at room temperature) 
was found to be strongly temperature dependent 
with a very high activation energy $\simeq 14000 $ K.  
Self diffusion has thus to be discarded as the origin of the break observed in $T_{1}$.
Previous neutron studies~\cite{Bee} have suggested that two types
of motion occur in the rotator phase of Cla:
tumbling of the molecular $C_{3}$ axis among six fourfold
crystallographic axes along the [100] directions,
and fast uniaxial rotation about the threefold axis.
It was shown that the motion around the molecular axes is very fast
($\simeq 10^{-12}$ s at room temperature) with an activation energy of about 1240 K.
Therefore, it takes place outside the time window   
where the dynamical crossover is detected at $T_{x}$.
Dipolar tumbling motions are thus mainly involved 
in the evolution of the spin-lattice relaxation.
Two separate temperature domains are clearly identified:
below $T_{x}$, 
both NMR experimental proton and carbon data are well described 
by a Frenkel model assuming an Arrhenian behavior
of tumbling and uniaxial rotations.
Above $T_{x}$, 
it was not possible
to describe the evolution of $T_{1}$ using the previous
model. Hence, we expect that 
some hypotheses of this model in particular the finite number of equilibrium positions
become irrelevant at high temperatures. 

Anisotropy in the rotational motion can be precisely seen from
the spin lattice relaxation times of the three different protonated carbons.
Effectively the main relaxation process of the methylene and methine carbons
is the dipole-dipole interaction between the $\mathrm{^{13}C}$
nuclei and their attached protons (see Fig.~\ref{figure2}b).
Relaxation times for carbon were measured as a function of temperature
in the rotator phase. 
The 
ratio $T_{1}(\mathrm{C4})/T_{1}(\mathrm{C2})$ 
 of the relaxational times for the methylene (C2, C4)
is displayed in Fig.~\ref{figure2}b.
A strongly anisotropic motion is found below $T_{x}$
where the ratio of the carbon relaxational times  
is about 0.25. This latter value
can be well reproduced by the previous Frenkel
model. 
A drastic deviation from the anisotropic behaviour occurs at high temperatures
where a ratio value of 0.7 is reached.
Since free isotropic rotational diffusion processus would correspond to
identical relaxational times for the methylene,
this result reveals the influence of the residual crystalline field
on the rotational dynamics.
The experimental heat capacity obtained by Oguni et Al.~\cite{Oguni},
in the Cla plastic phase is reported in Fig.~\ref{figure2}c.
The evolution of $C_{p}$ shows an anomalous hump 
which was reported to resemble
somewhat to a Schottky anomaly.
Owing to the nature of the Cla molecule which is mainly rigid,
conformational changes have to be discarded for the origin of the Schottky anomaly.
Clearly, the hump covers the entire temperature range
where the dynamical crossover is seen in our NMR investigation.
Its origin 
will be discussed in the light of the MD simulation results.

\begin{figure}[h]
\centerline{
\includegraphics[width=8.5cm,angle=270]{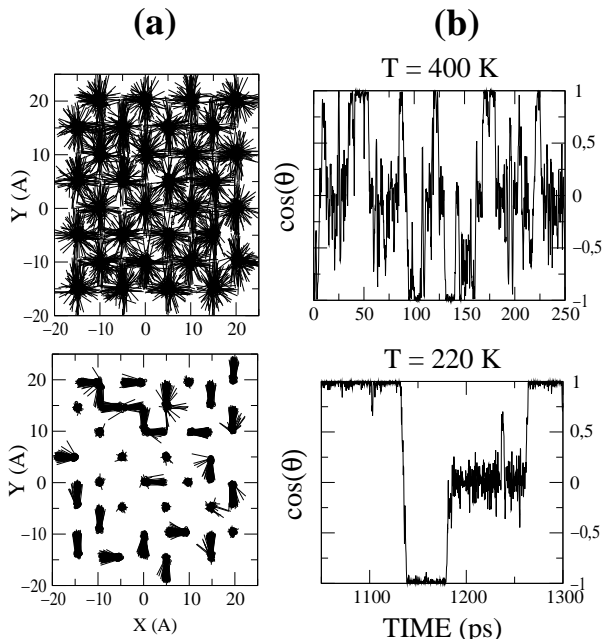}}
\caption{\protect\footnotesize
{\bf a}, Snapshots of the $\vec{\mu}$ dipolar moments
are projected
on one $xy$ plane at different instants
in a run of 250 ps above (400 K) and below (220 K)  $T_{x} \simeq 330 $ K.
At $T=$ 220 K, Some orientations are not
populated since the system is not completely equilibrated
{\it over this duration}.
{\bf b}, trajectory of the dipolar cosine angle $\theta$
of a target molecule at the same temperatures.
}
\label{figure1}
\end{figure}

MD simulations have been performed at
29 different temperatures from
$T =$ 220 to 500 K for
a sample corresponding to the crystalline fcc rotator phase
on a simple
model of Cla (see Fig.~\ref{figure1}).
A complete description of the model 
is given in~\cite{Affouard_AIP,model}. 
Owing to the very long MD runs, we succeeded in
investigating truly equilibrated states of the system
down to 220 K.
The reorientational motions can be described by a set of 
correlation functions which are defined as:
$ 
C_{l}(t)=
\frac{1}{N}\sum_{i=1,N}
\left \langle
P_{l} \left(
\vec{\mu}_{i}(t).\vec{\mu}_{i}(0)
\right)
\right \rangle
$
where $P_{l}$ is the $l$-order Legendre polynomial
and $\vec{\mu}_{i}$ is the individual dipolar moment of the molecule $i$.
The nature of dynamical changes and anisotropy
can be checked by computing $C_{l=1,2}$
functions which can also be related to 
the information obtained from NMR relaxation measurements.

\begin{figure}[h]
\centerline{\includegraphics[width=6.0cm,angle=270]{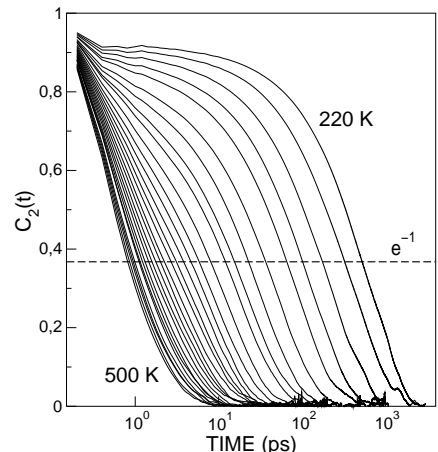}}
\caption{\protect\footnotesize
Orientational correlation function $C_{2}$
as function of time $t$
at different temperatures from 220 to 500 K.
}
\label{figure3}
\end{figure}

\begin{figure}[t]
\includegraphics[width=6.0cm]{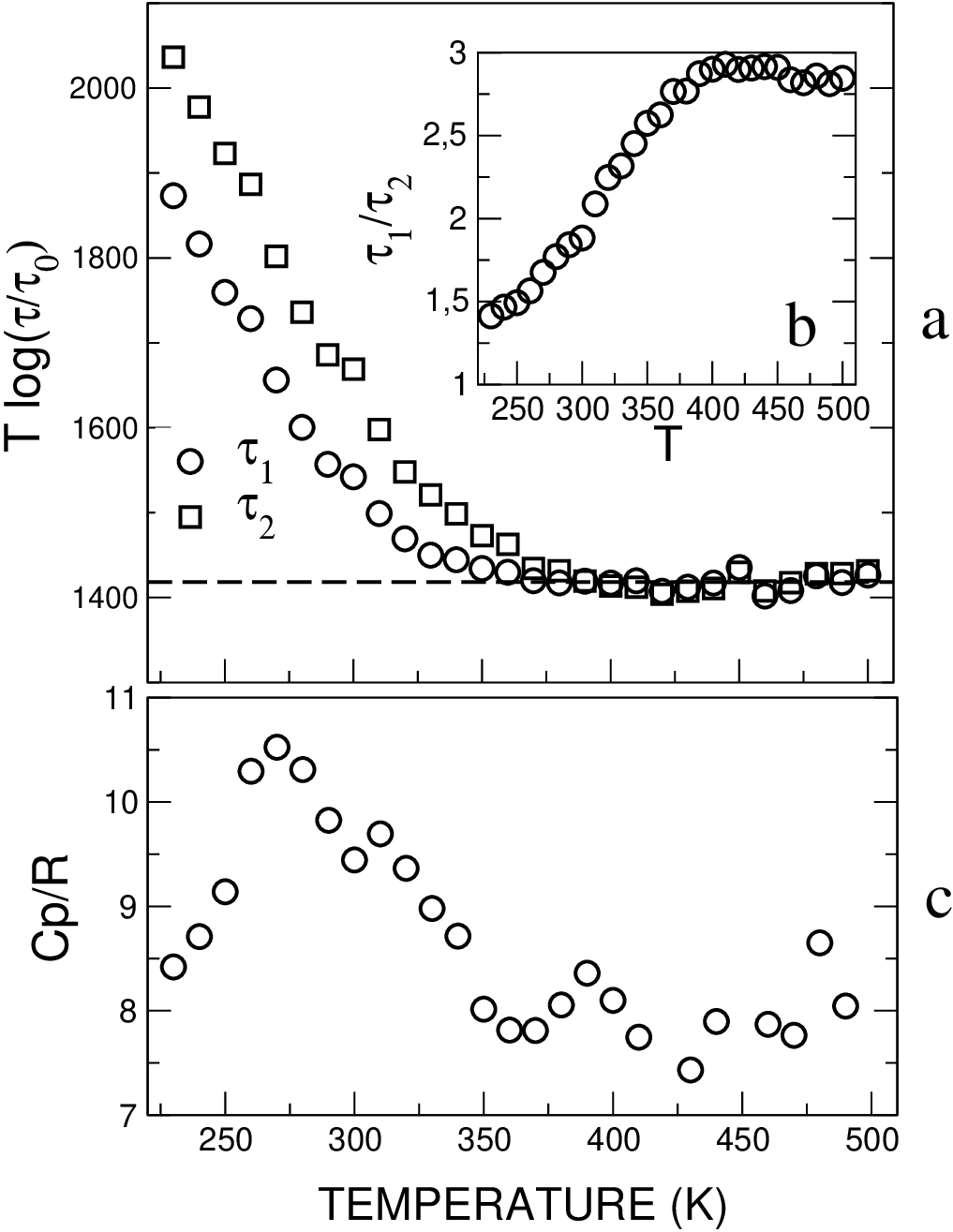}
\smallskip
\caption{\protect\footnotesize {\bf a}, orientational $\alpha$-relaxation time
$\tau_{1}$ (circle) and $\tau_{2}$ (square)
obtained by simulation as function of the temperature $T$.
The quantity plotted is $T\ln(\tau_{1,2}/\tau_{0})$
where the time $\tau_{0}$ is obtained by fitting the relaxation
times values for $T = [400-500] $ K to an Arrhenius form.
Deviation from the constant high-temperature value $\simeq 1418$ K
(dashed line) is observed
around $T_{x} \simeq 330 $ K. Low temperature apparent activation energy
is about 3109 K for $\tau_{1}$ and 3325 K for $\tau_{2}$ (solid line)
in good agreement with NMR experimental data.
{\bf b},
the ratio of the relaxational times $\tau_{1}/\tau_{2}$.
{\bf c}, calculated heat capacity as function of the temperature.}
\label{figure6}
\end{figure}

Fig.~\ref{figure3} shows the $C_{2}(t)$ time correlation
function for all investigated temperatures above 220 K.
Clearly,
a two step
relaxational behaviour classically
seen in supercooled molecular liquids~\cite{Kob,Lewis2}
is shown in our simulations.
When lowering the temperature,
an intermediate plateau region emerges
which proves
the presence of an orientational caging between
neighboring molecules.
This is the rotational
analogue of the translational \emph{cage effect} observed in liquids.
This transient regime is
followed by a slow
$\alpha$ process which can be associated in our system,
as checked for CNa~\cite{Affouard},
with large tumbling motion between fourfold crystallographic directions
where dipoles are preferably localized (see Fig.~\ref{figure1}).
A consequence is that the tumbling of one molecule
is allowed only if an orientational
rearrangement of its local neighbors occurs.
This \emph{cooperative
motion} is clearly displayed
using projections of the individual dipolar moments on one
crystallographic plane
in Fig.~\ref{figure1}.
The $\tau_{1}$ and $\tau_{2}$ relaxational times are defined
as the time it takes for their respective time correlation functions
to decay $e^{-1}$ of their initial values.
They are 
displayed in Fig.~\ref{figure6}a. 
Clearly, close to $T_{x} \simeq 330 $ K,
both $\tau_{1}$ and $\tau_{2}$ relaxational times start
diverging simultenously from the Arrhenian high temperature behaviour.
Both high and low temperature apparent
activation energies are found to be in good
agreement with NMR experimental results.
The nature of the rotational motions involved
in this dynamical change is given
by the ratio  $\tau_{1}/\tau_{2}$ in Fig.~\ref{figure6}b. 
This latter evolves from a value equal to 3 corresponding 
to free small-step rotational diffusion 
to a value of 1 classically  
associated to activated jump-like motion 
(see also Fig.~\ref{figure1})~\cite{Sindzingre,Sciortino}. 
These results prove for the first time the existence of a plastic-plastic 
transition which can be interpreted as the rotational analogue
of the Goldstein 
crossing temperature in liquids.
It could also be associated with a different process
of exploring the energy landscape topography
above and below $T_{x}$.
As it has been recently~\cite{Sastry,Glotzer,Heuer} proposed,  
dynamics of the system can be viewed as hopping processes 
between separated inherent structures at sufficiently low temperatures.
Specific heats have been directly extracted from the
calculated enthalpy ($C_{p} = dH/dT$) and are shown in Fig.~\ref{figure6}c.
Clearly, a pronounced hump is seen which extends over the low temperature
part of the rotator phase below 350 K. It proves that our MD simulation
captures the main features which are responsible to the anomaly
found in calorimetric experiments by Oguni et Al.~\cite{Oguni}.
A good agreement 
with the experimental results is found for
the area under
the $C_{p}$ hump and its 
temperature extension while for 
the position of the maximum it is only fair.
Obviously, the underlying absolute value
of the specific heat is not reproduced in this computation
owing to the simplicity
of our model, where
the internal vibrational contributions of the Cla molecule
and the fast uniaxial rotations have not been taken into account.
Thermodynamically, the anomalous hump requires an abrupt decrease
and a sharp change in curvature of the entropy vs temperature evolution.
Fig.~\ref{figure6}c shows that as the temperature is lowered
the $C_{p}$ hump begins to develop 
when dynamics change in nature.
The strong localization of the dipole along the [100] directions (see Fig.~\ref{figure1})
which develops below $T_{x}$, is thus the prime suspect
for the drastic change of the configurational entropy.
In conclusion, convincing experimental and numerical evidence
for a dynamical crossover ($T_{x}$) in the rotator phase 
of chloroadamantane have been obtained, very much in keeping
with recent views on structural glass-formers.
It is seen as a transition from quasi-free rotational diffusion 
of the molecular dipoles to activated geared tumblings at low temperatures.
This rotational localization is shown numerically to give rise to a 
specific heat hump as was recently measured~\cite{Oguni}.
The possible existence of such a calorimetric signature of a
dynamical accident in liquids is highly debated and was 
unsuccessfully searched until now~\cite{Murthy,Suga2}.
Certainly,
the underlying lattice and specificity of the orientational
disorder 
contribute to make dynamical changes
in the system 
much sharper than in liquids.
We now feel confident that orientationally disordered crystals provide
a valuable analog and alternative to
the conventional liquid glass-formers.
Results of the present study offer new interesting
possibilities of testing the different theoretical
approaches of the glass formation.

The authors are indebted to Professor K. Ngai for 
stimulating discussions.
This work was supported by the INTERREG II
program
(Nord Pas de Calais/Kent).
The authors wish to acknowledge the use of the facilities of the IDRIS
where some of the simulations were carried out.

\end{multicols}
\end{document}